\documentclass[10pt]{scrartcl}

\usepackage[english]{babel} 
\usepackage{lmodern}
\usepackage{amsmath}
\usepackage{hyperref}
\usepackage[all]{hypcap} 
\usepackage{amssymb}
\usepackage{cleveref}
\usepackage{bbm}
\usepackage{mathrsfs}
\usepackage{wasysym}
\usepackage{graphicx}
\usepackage{amsthm}
\usepackage{natbib}

\let\phi\varphi
\let\rho\varrho
\let\epsilon\varepsilon
\let\theta\vartheta
\newcommand{\dd}{\mathrm{d}}
\newcommand{\RR}{\mathbbm{R}}
\newcommand{\PP}{\mathbbm{P}}
\newcommand{\po}{{p_0}}
\newcommand{\thi}{\mathrm{th}}

\newtheorem{thm}{Theorem}

\newtheorem{Exam}[thm]{Example}
\newtheorem{rmk}[thm]{Remark}

\title{Generalizations of Mat{\'e}rn's\\ hard-core point processes}

\author{\small J. Teichmann $^a$, F. Ballani $^a$, K. G. van den Boogaart $^b$}

\date{\small \today}

\begin{document}

\maketitle

\begin{center}
\small $^a$\,Institut f\"ur Stochastik, Technische Universit\"at Bergakademie Freiberg,\\
Pr\"uferstr. 9, D-09596 Freiberg, Germany, \par \texttt{jakob.teichmann@math.tu-freiberg.de} / \texttt{ballani@math.tu-freiberg.de}  

$^b$\,Helmholtz Institute Freiberg for Resource Technology, TU Bergakademie Freiberg,\\
Halsbr\"ucker Str. 34, D-09599 Freiberg, Germany, \par \texttt{boogaart@math.tu-freiberg.de}
\end{center}


\textbf{Abstract.} Mat{\'e}rn's hard-core processes are valuable point process models in spatial statistics. In order to extend their field of application, Mat{\'e}rn's original models are generalized here, both as point processes and particle processes. The thinning rule uses a distance-dependent probability function, which controls deletion of points close together. For this general setting, explicit formulas for first- and second-order characteristics can be given. Two examples from materials science illustrate the application of the models.\\ \par

\textbf{Key words:} Point process, marked Poisson process, Mat{\'e}rn hard-core process, dependent thinning, independent thinning, pair correlation function.

%
\section{Introduction}
The present paper aims to generalize Mat{\'e}rn's well-known first and second hard-core point processes. These point process models, introduced by B.~Mat{\'e}rn \citep{M60,M}, are typical examples for models derived from Poisson point processes, the latter being an important basis for constructing more complicated point processes, random sets and fibre processes at all \citep{SK}. They have been successfully applied to real data, for instance, they have been used in ecological \citep[e.\,g.][]{PKD,S4,W} and CSMA network modeling \citep[e.\,g.][]{BB,BC,H1} as well as geographical analysis \citep{S3}. 

Mat{\'e}rn's first and second hard-core point processes are derived by applying a specific thinning rule to a homogeneous Poisson point process in $\RR^d$. As such they are important examples for \emph{dependent thinning} \citep[][]{I,S3} where the thinning depends on the underlying process, somehow. For instance, the \emph{Mat{\'e}rn I} hard-core point process is obtained by deleting every point in the process with its nearest neighbor closer than a given hard-core distance \citep[cf.][]{M60,M}. In general, a thinning operation or rule \citep{I} determines which points in the basic process are deleted. For example, such thinnings drive the evolution of plant communities due to competition-induced mortality \citep{BM}. 

In contrast to depend thinning, the well-known $p$- and $p(x)$-\emph{thinning} approaches described in \citep[Section 6.2.1]{I} or \citep[Section 11.3]{D1} use \emph{independent thinning}. That means that the thinning operation is independent of the configuration of the underlying point process and, at position $x$, a point will be deleted with some deterministic probability $1-p$ or $1-p(x)$, respectively. 

In the following, the way of thinning a Poisson point process in order to obtain a Mat{\'e}rn hard-core point process is generalized in two directions. The first idea is to combine both independent and dependent thinning which can simply be interpreted as a subsequent independent thinning of a dependently thinned point process. The second idea is that in the dependent thinning a distance-dependent probability function $f$ controls deletion of points which are close together. This means that, depending on the distance to its neighbors, a point will be deleted only with some probability and not surely as is the case, e.\,g., in the Mat{\'e}rn I thinning rule. Thus it seems to be justified to speak of a \emph{probabilistic thinning rule} in what follows.
 
An application of appropriate probabilistic thinning rules to Poisson point processes then leads to a class of point processes which are generalizations of the Mat{\'e}rn hard-core point processes. As distinguished from Gibbs point processes \citep{SK,I,D1}, explicit formulas for first- and second-order characteristics can then still be derived. By means of several examples for the probability function $f$ it is shown that soft-core, hard-core as well as aggregative point processes can be obtained by this approach which reveals its high flexibility.

This paper is organized as follows. In \Cref{sec2}, first the generalization of the Mat{\'e}rn I hard-core point process based on the thinning of a homogeneous Poisson process is introduced and first-order and second-order characteristics are given. Based on the idea of \citet{S2} and \citet{Ma} to extend the original Mat{\'e}rn II hard-core model by giving each point a random radius, where in \citep{Ma} as a special case also a respective extension of the Mat{\'e}rn I process is covered, in \Cref{modelmark}, the dependent thinning model is further enhanced to marked point processes which is useful to model particle systems. \Cref{modelgeneral} contains then a discussion of a related generalization of Mat{\'e}rn-II-type point processes. Finally, in \Cref{application}, the applicability of the new models is illustrated by means of two data sets. 
%
%
\section{Probabilistic thinning model -  Mat{\'e}rn I case}\label{sec2}
\subsection{Model description}\label{model}
Let $\Phi$ be a homogeneous Poisson point process in $\RR^d$ with intensity $\lambda>0$ \citep[see][]{D} on a probability space $(\Omega,\mathcal{F},\PP)$, $\po\,\in\,]0,1]$, and $f:[0,\infty[ \longrightarrow [0,1]$ be a measurable function. Denote by $\|\cdot\|$ Euclidean distance in $\RR^d$. From $\Phi$ a new model $\Phi_{\thi}$ is derived by applying the following \emph{probabilistic dependent thinning} rule to $\Phi$. A point $x\in\Phi$ is retained with probability
\begin{equation}\label{eqn:rule1}
p(x,\Phi)=\po\prod_{\substack{y\in \Phi\\ y\neq x}}[1-f(\|x-y\|)]\,
\end{equation}
independently from deleting or retaining other points of $\Phi$.
This means that two points a distance $r>0$ apart delete each other independently with probability $f(r)$. Independently from deleting due to pairwise interaction, each (surviving) point is (then) deleted with probability $1-\po$.

Since the homogeneous Poisson process $\Phi$ is both stationary and isotropic and the thinning rule is independent both from location and direction, the thinned point process $\Phi_{\thi}$ is stationary and isotropic as well.

In the following we will sometimes write $\mathrm{MatI}[\lambda,p_0,f]$ for the distribution of $\Phi_{\thi}$.

\begin{rmk}
Taking $\po=1$ and $f=\mathbbm{1}_{[0,R]}$ for some $R>0$, the point process $\Phi_{\thi}$ coincides with the Mat{\'e}rn I hard-core process with hard-core distance $R$ since then any two points a distance $r\leq R$ apart delete each other with probability 1, i.\,e. almost surely.
\end{rmk}

\begin{rmk}
The thinning rule (\ref{eqn:rule1}) could be refined by making the retention probability dependent also on three-point or further multi-point configurations, or, more general, on any functional which depends both on the point $x\in\Phi$ and the point pattern $\Phi$. An example for the latter would be the number $n_R(x,\Phi)$ of neighbors of $x\in\Phi$ within a certain distance $R>0$ which is, e.\,g., directly used in the definition of Strauss point processes \citep{Str}. In fact, even the choice $f=f_0\mathbbm{1}_{[0,R]}$ for some $f_0\in\,]0,1[$ results in a retention probability
\begin{equation*}
p(x,\Phi)=p_0\,(1-f_0)^{n_R(x,\Phi)}.
\end{equation*}
\end{rmk}
%
%
\subsection{First- and second-order characteristics}\label{pcf}
As is known for the original Mat{\'e}rn I hard-core point process, explicit formulas both for its intensity and its pair correlation function \citep{D,I,Mo,SK} can be stated \citep{D,I,M60,M,S2}. Although the definition of the thinned point process $\Phi_{\thi}$ is more complicated, arguments similar to that given for instance in \citep{D} can be used to derive the subsequent expressions in \Cref{eq9,eq15} for both the intensity and the pair correlation function of $\Phi_{\thi}$. The respective proofs are omitted here due to the fact that the model $\mathrm{MatI}[\lambda,\po,f]$ also appears as a special case of another model introduced later in \Cref{modelmark}.
\begin{thm}[Intensity]\label[theorem]{eq9}
The intensity $\lambda_{\thi}$ of the thinned point process $\Phi_{\thi}$ is
\begin{align*}
\lambda_{\thi}= \lambda\,\po\,\exp\left(-\lambda d\,b_d\int_0^\infty f(r)r^{d-1} \mathrm{d} r\right),
\end{align*}
where $b_d$ denotes the volume of the unit ball in $\RR^d$.
\end{thm}
In case the integral in \Cref{eq9} is infinite, the resulting intensity $\lambda_{\thi}$ vanishes. That is, $\Phi_{\thi}$ contains almost surely no points since the applied thinning is so strong that all points of $\Phi$ are removed almost surely. Since this case is uninteresting, in what follows we thus consider only those functions $f$ which satisfy the integrability condition
\begin{align}\label{integrab}
\int_0^\infty r^{d-1}f(r)\mathrm{d} r < \infty.
\end{align}
Let $f \astrosun f$ denote the \emph{radial self-convolution} of $f$, i.\,e. 
\begin{equation}\label{eqn:selfconv}
[f \astrosun f](r)=\int_{\RR^d} f(\|x\|)f(\|x-r\cdot v\|) \mathrm{d} x
\end{equation}
for $r\geq 0$ and $v\in \RR^d$ with $\|v\|=1$, which is the $d$-dimensional \emph{convolution} of $f(\|\cdot\|)$ with itself at point $r\cdot v$. 
\begin{thm}[Pair correlation function]\label[theorem]{eq15}
The pair correlation function $g_{\thi}$ of the thinned point process $\Phi_{\thi}$ is
\begin{align*}
g_{\thi}=(1-f)^2\exp\left(\lambda\, f \astrosun f\right). 
\end{align*}
\end{thm}
The following two examples illustrate for dimension $d=2$ how certain choices of $f$ influence the second-order behavior of the resulting thinned point processes.
\begin{Exam}\label[example]{ex:1}
Let $\lambda>0$ be arbitrary and $\po=1$. Consider $\mathrm{MatI}[\lambda,1,f_{a,R}]$ with
\begin{align*}
f_{a,R}(r)=
\begin{cases}
    1 & 0 \leq r \leq a\\
    \exp\left(-\frac{r^2-a^2}{R^2-a^2}\right) & \text{otherwise}
\end{cases}
\end{align*}
for $R>0$ and $a\in\,[0,R]$\,. Then it is plain to see that for any $\lambda$ and $R$ fixed the intensity of $\Phi_{\thi}$, which is distributed according to $\mathrm{MatI}[\lambda,1,f_{a,R}]$, is
\begin{align*}
\lambda_{\thi}=\lambda \exp(-\pi\lambda R^2),
\end{align*} 
i.\,e., it does not depend on $a$. However, depending on the parameter $a$, the pair correlation function of $\Phi_{\thi}$ shows a certain range of second-order behavior, see \Cref{example1}.
\begin{figure}
\begin{center}
\includegraphics[width=5.5cm]{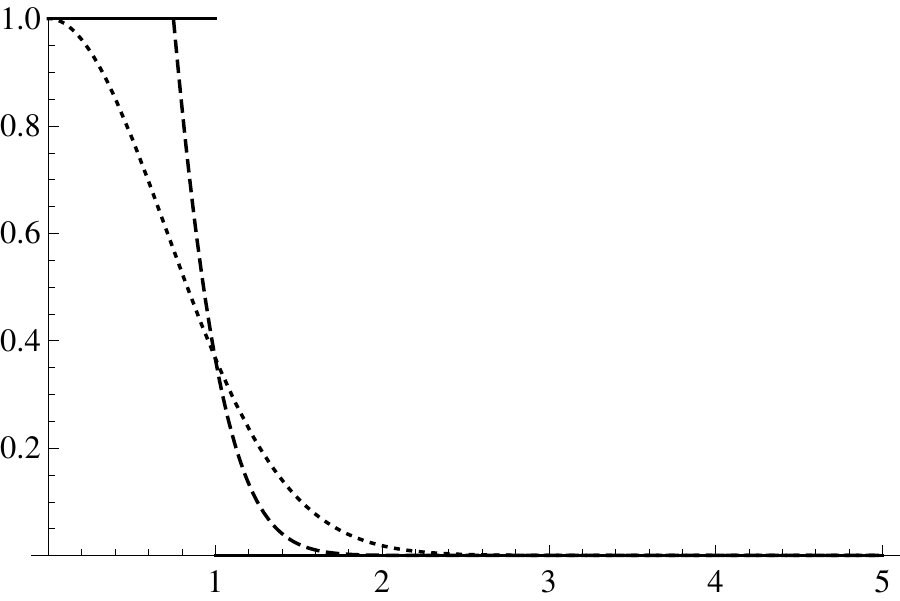}
\hspace{15mm}
\includegraphics[width=5.5cm]{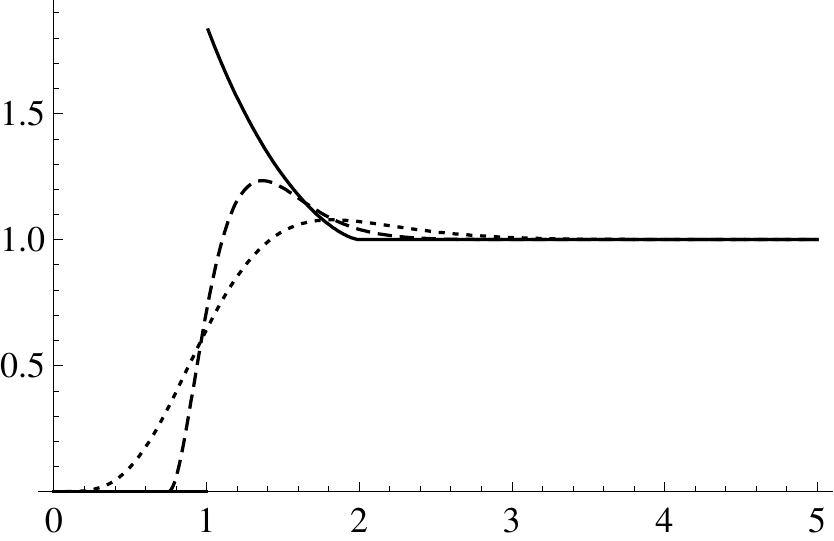}
\caption{Left, plot of $f_{a,1}$ for $a=0$ (dotted), $a=0.75$ (dashed) and $a=1$ (solid). Right, corresponding pair correlation functions for $\lambda=1/2$.}
\label[figure]{example1}
\end{center}
\end{figure}
For $a=0$ the thinning generates a pure \emph{soft-core} point process, i.\,e. thinning is the stronger the closer point pairs of the initial Poisson process are but each pair distance has still non-vanishing probability. In the other direction, the Mat{\'e}rn I hard-core model is included as the limit for $a\to R$. A mixture of both hard- and soft-core type behavior can be achieved with $a \in\,]0,1[$.
\end{Exam}
\begin{Exam}\label[example]{ex:2}
Another type of family $\mathrm{MatI}[\lambda,1,h_a]$, where again $\lambda>0$ is arbitrary and $\po=1$, is given by setting
\begin{align*}
h_a(r)=\frac{r^a\,\exp(-r^2)}{\Gamma\left(1+\frac{a}{2}\right)}\,,\quad r\geq 0
\end{align*}
for $a\in\,[0,\infty[$, where $\Gamma$ denotes the $\Gamma$-function. Again, for $\lambda$ fixed, all resulting thinned point processes have the same intensity
\begin{align*}
\lambda_{\thi}=\lambda \exp(-\pi\lambda),
\end{align*} 
i.\,e. independent from $a\geq 0$. Here, the thinning based on $h_a$ results in \emph{aggregation} or \emph{cluster-like} processes which is indicated by pair correlation functions with its maximum in the origin, see \Cref{example2}, since points close together are deleted with relatively low probability. 
\begin{figure}
\begin{center}
\includegraphics[width=5.5cm]{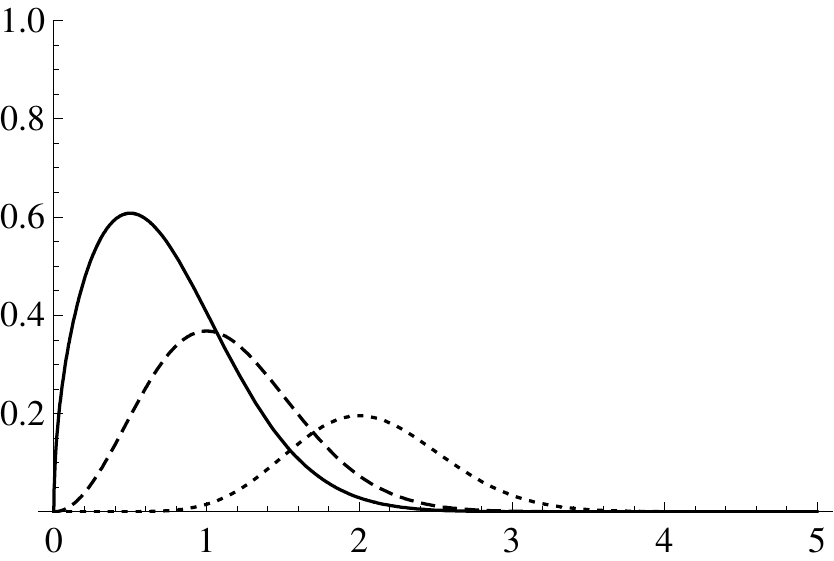}
\hspace{15mm}
\includegraphics[width=5.5cm]{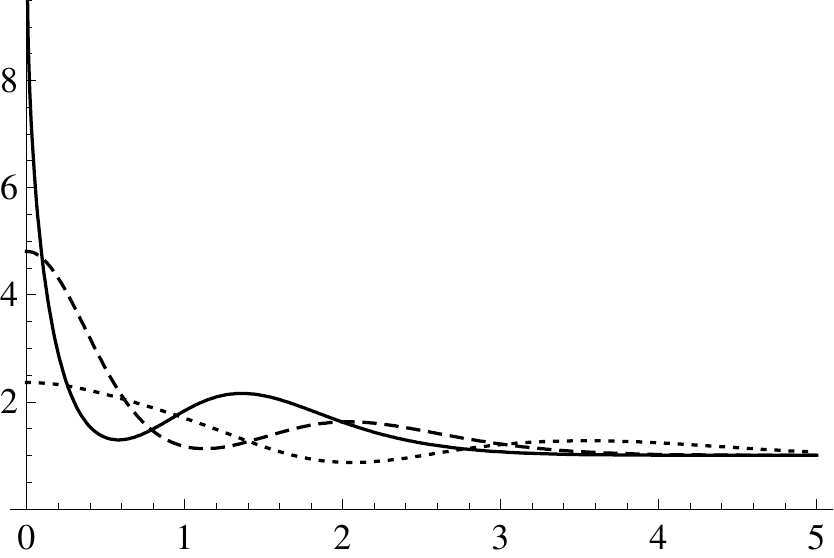}
\caption{Left, plot of $f_{a}$ for $a=8$ (dotted), $a=2$ (dashed) and $a=0.5$ (solid). Right, corresponding pair correlation functions for $\lambda=2$.}
\label[figure]{example2}
\end{center}
\end{figure}
\end{Exam}
\Cref{ex:1,ex:2} show that the behavior of the pair correlation function of the thinned process $\Phi_{\thi}$ might be adjusted using appropriate functions $f$ such that it is possible to model soft-core, hard-core as well as aggregative point processes \citep[cf.][]{I}. Furthermore, a 'mixture' of these types can be obtained combining the corresponding functions $f$. In summary, this reveals high flexibility of this approach with respect to second-order properties.
%
%
\section{Probabilistic thinning for marked Poisson processes}\label{modelmark}
Marked point processes as generalizations of usual point processes are highly relevant in practical applications. Many ecological and environmental systems can be described by marked point processes \citep[see][]{Ga}. Furthermore, there is a large literature on processes of non-overlapping grains in physics and chemistry \citep[see][]{A,Ma}. Here, Mat{\'e}rn hard-core processes are equipped with random radii as marks. While in the previous section a dependent thinning model generalizing the usual Mat{\'e}rn I hard-core point process was introduced, the aim of the present section is to carry over this approach to respective marked point processes. Again, intensity and pair correlation function of the corresponding unmarked point process can be given explicitly.

Let $\Psi$ be a homogeneous independently marked Poisson point process in $\RR^d$ with intensity $\lambda$ and independent and identically distributed (i.\,i.\,d.) real-valued marks with $\mu$ as its mark distribution. Furthermore, let $\po\in\,]0,1]$, and $f:[0,\infty[ \times \RR^2 \longrightarrow [0,1]$ be a fixed measurable function satisfying $f(\cdot,m,n)=f(\cdot,n,m)$ for all $m,n \in \RR$. 

From $\Psi$ a new model $\Psi_{\thi}$ is derived by applying the following probabilistic dependent thinning rule to $\Psi$. The marked point $(x,m)\in \Psi$ is retained with probability
\begin{equation}\label{eqn:rule2}
p(x,m,\Psi)=\po\prod_{\substack{(y,n)\in\Psi\\ y\neq x}}\big[1-f(\|x-y\|,m,n)\big]
\end{equation}
independently from deleting or retaining other marked points of $\Psi$.
This means that two points of distance $r>0$ apart with marks $m$ and $n$ delete each other independently with probability $f(r,m,n)$, and, again, independently from deleting due to pairwise interaction, each surviving point is then additionally deleted with probability $1-\po$.

Since the thinning rule is again independent both from location and direction, the point process $\tilde{\Psi}_{\thi}$ of unmarked points of $\Psi_{\thi}$ inherits both stationarity and isotropy from the homogeneous Poisson process of unmarked points of $\Psi$.

In the following we will sometimes write $\mathrm{MatI}[\lambda,\mu,p_0,f]$ for the distribution of $\Psi_{\thi}$. Of course, if $f$ does not depend on the marks the unmarked point process $\tilde{\Psi}_{\thi}$ coincides with the model $\mathrm{MatI}[\lambda,p_0,f]$ introduced in \Cref{model}, and all formulas given there appear as particular cases of the subsequent results.   
\begin{Exam}
Taking $f(r,m,n)=\mathbbm{1}_{[0,m+n]}(r)$, $r,m,n\geq 0$, and random marks  uniformly from $[0,1]$, we obtain a hard-core point process with random hard-core radii, i.\,e., we can interpret the retained marked point $(x,m)$ as a ball of radius $m$ centered in $x$, see \Cref{radii}. This example was studied by \citet{Ma} as a model for systems of varying-sized, non-overlapping spherical grains. 
\begin{figure}
\begin{center}
\includegraphics[width=5.5cm]{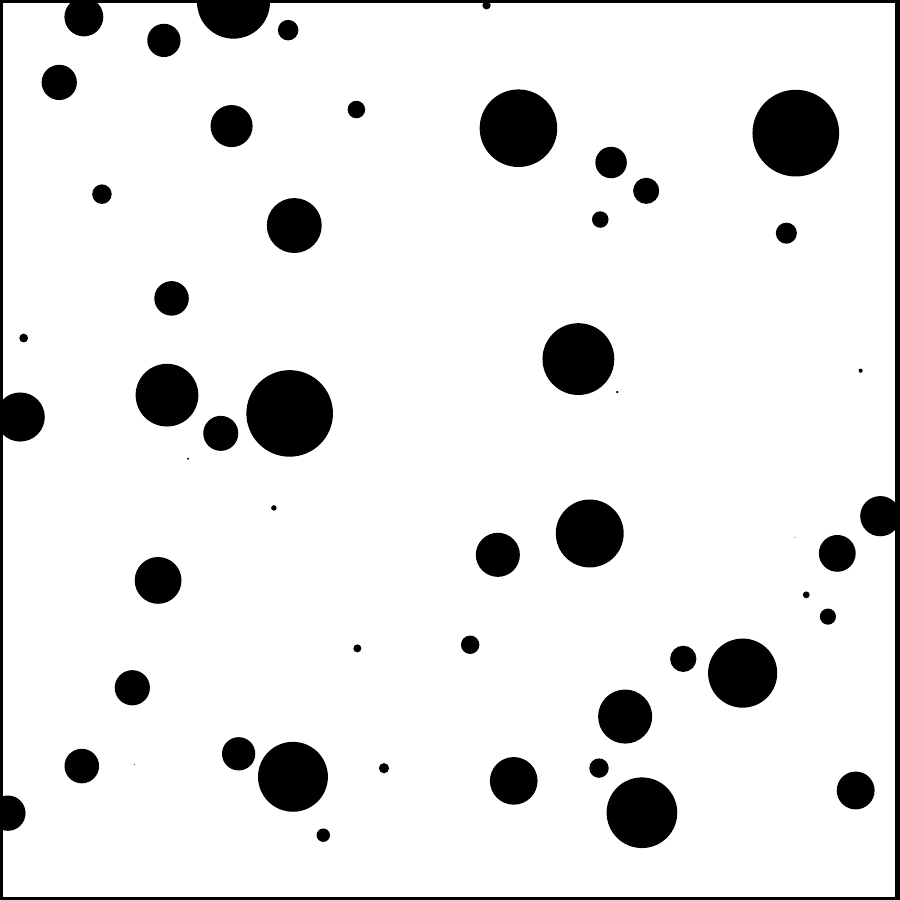}
\caption{Sample of $\Psi_{\thi}$ in dimension $d=2$ for $\lambda=100$, $\po=1$, $f(\cdot,m,n)= \mathbbm{1}_{[0,m+n]}$ and $\mu$ the uniform distribution on $[0,1]$.}
\label[figure]{radii}
\end{center}
\end{figure}
\end{Exam}
\begin{thm}[Intensity]
The intensity $\lambda_{\thi}$ of the thinned point process $\tilde{\Psi}_{\thi}$ is
\begin{align*}
\lambda_{\thi}= \lambda\,\po\int_{\RR} \exp\left(-\lambda d\,b_d \int_{\RR}  \int_0^\infty f(r,m,l)r^{d-1} \mathrm{d} r \, \mu(\dd l) \right) \mu(\dd m) .
\end{align*}
\label[theorem]{eq25}
\end{thm} 
As an example, consider once more the case $\po=1$ and $f(\cdot,m,l)= \mathbbm{1}_{[0,m+l]}$ with positive-valued mark distribution $\mu$. Then the result in \Cref{eq25} reads
\begin{align*}
\lambda_{\thi}= \lambda \int_0^\infty \exp\left(-\lambda b_d \int_0^\infty  (l+m)^d \, \mu(\dd l) \right) \mu(\dd m) \,,
\end{align*}
which coincides with the formula in \citep[Section 2.2, Example 2.1]{Ma}. 
\begin{thm}[Pair correlation function]\label[theorem]{eq31}
The pair correlation function of the thinned point process $\tilde{\Psi}_{\thi}$ is 
\begin{align}
g_{\thi}(r)=\frac{\lambda^2\,\po^2}{\lambda_{\thi}^2}\int_\RR\int_\RR (1-f(r,m,n))^2\,q_m\,q_n\,q_{m,n}(r)\,\mu(\dd m)\, \mu(\dd n),\quad r\geq 0,
\end{align}
where
\begin{align}\label{pcfradii}
q_{m,n}(r)=\exp\left(\lambda\,\int_{\RR} f(\cdot,m,l) \astrosun f(\cdot ,n,l)(r) \,\mu(\dd l)\right),\quad r\geq 0
\end{align}
and
\begin{align}\label{intensradii}
q_m=\exp\left(-\lambda\,\int_{\RR}\int_{\RR^d} f(\|x\|,m,l)\,\dd x\,\mu(\dd l)\right).
\end{align}
\end{thm}
Note that $g_{\thi}(r)$ actually does not depend on $\po$ since it appears also as a factor in $\lambda_{\thi}$ and thus cancels out.
%
%
\section{Probabilistic thinning model - Mat{\'e}rn II case}\label{modelgeneral}
In the previous sections, a certain kind of dependent thinning has been introduced where two competing points of an underlying Poisson process are \emph{both} removed with some probability depending on a deterministic function $f$. While this is a generalization of the classical Mat{\'e}rn I hard-core point process, in the present section we aim to generalize the classical Mat{\'e}rn II model in a similar fashion. For that purpose, weights will be assigned once to all the points and in a competition between two points only \emph{one} of them, namely the point equipped with weight greater than or equal to the weight of the other point will be removed with some probability. Again, we can derive expressions for both the intensity and the pair correlation function of the resulting process of unmarked points.\\

Let $\Pi$ be a homogeneous independently marked Poisson point process in $\RR^d$ with intensity $\lambda$ and i.\,i.\,d. $\RR^2$-valued marks. The first component $m$ of any such random bivariate mark $(m,v)$ has distribution $\mu$ and the second component $v$ has distribution $\nu_m$ which might depend on $m$. The mark $m$ plays the same role as in \Cref{modelmark} and is sometimes referred to as 'mark' whereas $v$ serves as a weight used in the thinning procedure and is thus sometimes referred to as 'weight'.  Furthermore, let again $\po\,\in\,]0,1]$, and $f:[0,\infty[ \,\times\, \RR^2 \longrightarrow [0,1]$ be a fixed measurable mapping satisfying $f(\cdot,m,n)=f(\cdot,n,m)$ for all $m,n \in \RR$. 

From $\Pi$ a new model $\Pi_{\thi}$ is derived by applying the following \emph{probabilistic dependent thinning} rule. The point $(x,m,v)\in\Pi$ is kept as a point of $\Pi_{\thi}$ with probability
\begin{equation}\label{eqn:rule3}
p(x,m,v,\Pi)=\po\prod_{\substack{(y,n,w)\in \Pi\\ y\neq x}}\big[1-\mathbbm{1}\{v\geq w\}\,f(\|x-y\|,m,n)\big]
\end{equation}
independently from deleting or retaining other points of $\Pi$, where $\mathbbm{1}\{A\}$ is the indicator of event $A$.
This means that if two points with marks $m$ and $n$ are a distance $r>0$ apart then only the point with weight greater than or equal to the weight of the other point is deleted by the other point with probability $f(r,m,n)$. Additionally, each surviving point is then again independently $\po$-thinned. 

In the following we will denote by $\mathrm{MatII}[\lambda,\mu,(\nu_m)_{m\in \RR},p_0,f]$ the distribution of $\Pi_{\thi}$.

Note that the meaning of the weights in the thinning rule (\ref{eqn:rule3}) is here in accordance with the meaning of the respective weights used in the definition of the original Mat{\'e}rn II hard-core point process in most of the literature \citep{I,SK,S2} (but not \citep{Ma}), i.\,e., they have to be understood more (biologically) as \emph{times of appearance} than \emph{importance weights}, and in a competition the lower weight wins.
\begin{rmk}\label[remark]{rmk:MIIspheres}
Taking $f(\cdot,m,n)= \mathbbm{1}_{[0,m+n]}(\cdot)$, $m,n\geq 0$, and marks according to a positive mark distribution $\mu$ results in a hard-core process with random hard-core radii. This case was also studied by \citet{Ma} as an extension of Mat{\'e}rn's second hard-core point process to random configurations of non-overlapping spheres. The original Mat{\'e}rn II point process with hard-core radius $R>0$ can be obtained using $\mu=\delta_R$ and $\nu_m$ as the uniform distribution on $[0,1]$.
\end{rmk}
\begin{rmk}
The particular choice $\nu_m=\delta_1$ for all marks $m$ leads back to the model $\Psi_{\thi}$ described in \Cref{modelmark}. Here, all points would have the same weight such that two competing points delete each other independently with the same probability. 
\end{rmk}
Denote by $F_{\tau}$ the cumulative distribution function of a probability measure $\tau$ on $\RR$, i.\,e. $F_{\tau}(t)=\tau(]-\infty,t])$, $t\in\RR$.

For simplicity assume in what follows that the weight distributions $\nu_l$ are all (absolutely) continuous (w.r.t. lebesgue measure) but may still depend on mark $l$. The main effect is that the event $\mathbbm{1}\{v\geq w\}$ has probability zero and hence only one of two competing points is deleted with some probability. In particular, this excludes the Mat{\'e}rn-I-like processes introduced in the previous sections.

Furthermore, let us recall that the point process $\tilde{\Pi}_{\thi}$ of unmarked points of $\Pi_{\thi}$ inherits both stationarity and isotropy from the homogeneous Poisson process of unmarked points of $\Pi$.
\begin{thm}[Intensity]\label[theorem]{eq43}
The intensity $\lambda_{\thi}$ of the thinned point process $\tilde{\Pi}_{\thi}$ is
\begin{align*}
\lambda_{\thi}= \lambda\,\po \int_\RR \int_\RR q_m(w)\,\nu_m(\dd w)\, \mu(\dd m)\,,
\end{align*}
where
\begin{equation}\label{intens:enh}
q_m(w)=\exp\left(-\lambda\,\int_{\RR} F_{\nu_l}(w)\int_{\RR^d} \,f(\|x\|,m,l)\, \dd x\,\mu(\dd l)\right).
\end{equation}
\end{thm}
In the special case that the weight distribution $\nu_l$ does not depend on the mark $l$, i.\,e., $\nu_l=\nu$ for some continuous distribution $\nu$, \Cref{eq43} simplifies to
\begin{align}
\lambda_{\thi}
&=\lambda\,\po\int_\RR\int_\RR\exp\left(-\lambda\,F_\nu(w)\int_{\RR}\int_{\RR^d} \,f(\|x\|,m,l) \dd x\,\mu(\dd l)\right)\nu(\dd w)\, \mu(\dd m)\notag \\
&=\lambda\,\po\int_\RR\int_0^1\exp\left(-\lambda\,u\int_{\RR}\int_{\RR^d} \,f(\|x\|,m,l) \dd x\,\mu(\dd l)\right)\,\dd u\, \mu(\dd m)\notag\\
&=\po\int_\RR\frac{1-\exp\left(-\lambda\,\int_{\RR}\int_{\RR^d} \,f(\|x\|,m,l) \dd x\,\mu(\dd l)\right)}{\int_{\RR}\int_{\RR^d} \,f(\|x\|,m,l) \dd x\,\mu(\dd l)}\, \mu(\dd m)\,,\label{eq44}
\end{align}
by change of variables. In the case $\po=1$ and $f(\cdot,m,n)=\mathbbm{1}_{[0,m+n]}$ formula (\ref{eq44}) coincides with the result stated in \citep[Theorem 3.1]{Ma}. 
\begin{rmk}\label[remark]{rmk:comparison}
>From \Cref{intens:enh} it is easy to see that, due to $F_{\nu_l}(w)\leq1$ for all $w$, the intensity of the thinned point process $\Pi_{\thi}$ is always greater than the intensity of a thinned point process according to $\mathrm{MatI}[\lambda,\mu,\po,f]$ from \Cref{modelmark} with the same parameters.
\end{rmk}

\begin{thm}[Pair correlation function]\label[theorem]{eq49}
The pair correlation function $g_{\thi}$ of the thinned point process $\tilde{\Pi}_{\thi}$ is
\begin{align*}
g_{\thi}(r)=& \frac{\lambda^2\,\po^2}{\lambda_{\thi}^2}\int_\RR \int_\RR (1-f(r,m,n))\,I_r(m,n)\,\mu(\dd m)\, \mu(\dd n)\,,\quad r>0,
\end{align*}
where
\begin{align*}
I_r(m,n)= \int_\RR \int_\RR q_m(w)\,q_n(t)\,q_{m,n}(w,t,r)\,\nu_m(\dd w)\,\nu_n(\dd t)
\end{align*}
with $q_m(w)$ from \Cref{intens:enh} and
\begin{align*}
q_{m,n}(w,t,r)=
\exp\bigg(\lambda\int_\RR F_{\nu_l}(\min \{w,t\})\,[f(\cdot,m,l) \astrosun f(\cdot ,n,l)](r) \,\mu(\dd l) \bigg).
\end{align*}
\end{thm}
Again $g_{\thi}(r)$ does not depend on $\po$. In the special case where $\nu_l=\nu$ does not depend on $l$, $I_r(m,n)$ can be written as
\begin{align*}
I_r(m,n)&= \int_0^1 \int_0^1 q_m^s\,q_n^t\,q_{m,n}(r)^{\min\{s,t\}}\,\dd s\,\dd t=J(a,b,c)+J(b,a,c)
\end{align*}
where
\begin{align*}
J(a,b,c)&=\frac{1}{b(a+b+c)}+\frac{1}{a+c}\,\bigg(\frac{\exp(a+b+c)}{a+b+c}-\frac{\exp(b)}{b}\bigg)
\end{align*}
with $q_m$ and $q_{m,n}(r)$ according to \Cref{intensradii} and \Cref{pcfradii}, respectively, and $a=\log q_m$, $b=\log q_n$ and $c=\log q_{m,n}(r)$.
%
%
\section{Applications}\label{application}
\subsection{Fontainebleau sandstone}\label{font}
\begin{figure}
\begin{center}
\includegraphics[width=5.5cm]{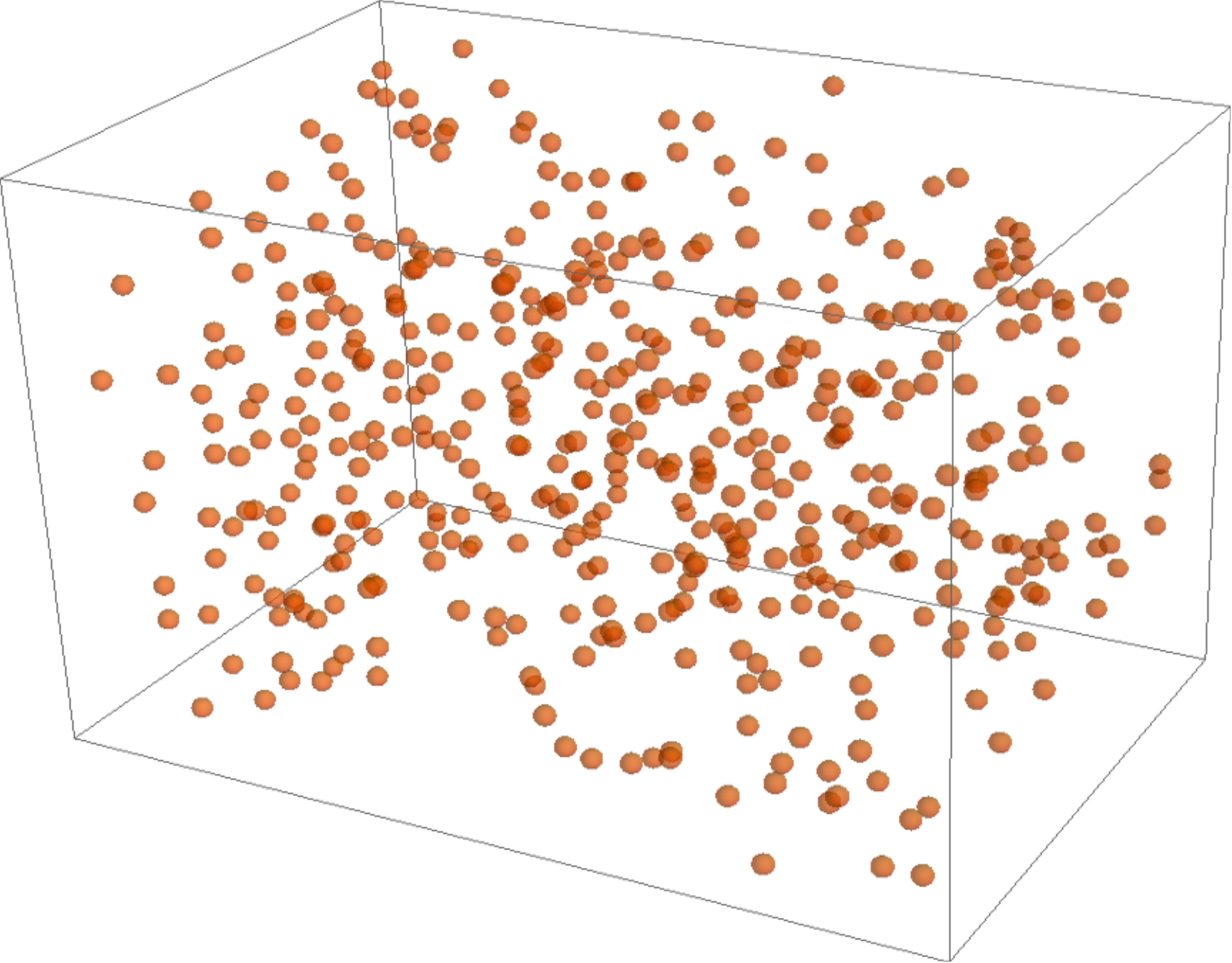}
\hspace{15mm}
\includegraphics[width=5.5cm]{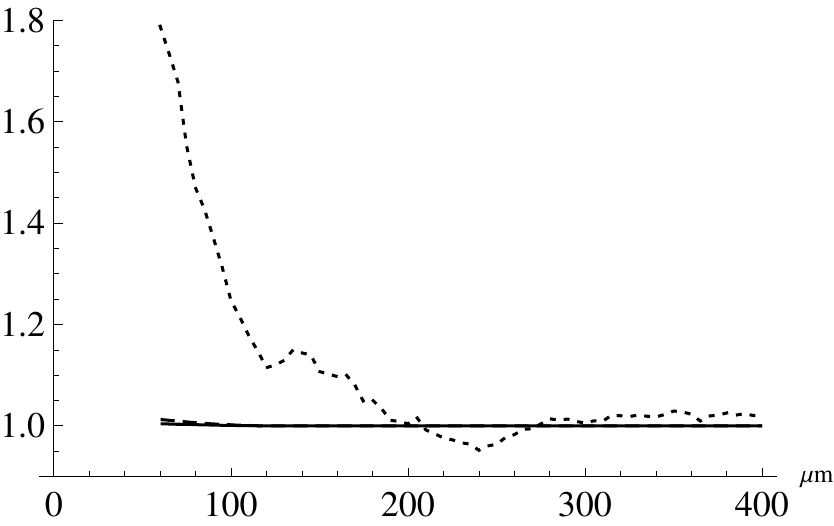}
\caption{Left, 470 nodes of a network adapted to the pores of a $2 \times 2 \times 3$mm sample of Fontainebleau sandstone. Right, the empirical pcf $\hat g$ (dotted), $g$ for Mat{\'e}rn I (dashed) and for Mat{\'e}rn II (solid line).}
\label[figure]{Materntest}
\end{center}
\end{figure}

The first data set is a point pattern describing the pore network of a sample of Fontainebleau sandstone. A visualization is given in \Cref{Materntest}. A detailed description how this point pattern was obtained can be found in \citep{So}. It has been further analyzed in the literature, for instance, \citet{Ts} discuss second-order characteristics and a certain Euler-Poincar{\'e} characteristic connected with the data. A standard test of the hypothesis that the pattern is of CSR type (complete spatial randomness) \citep{I} results in rejection with a $p$-value of $0.0002$.

The minimum interpoint distance in the pattern is $60.7\,\mu\mathrm{m}$, and it is just this hard-core distance which leads to a rejection of the CSR hypothesis. Consequently, a hard-core point process model seems to be more appropriate for this data. Because of the low point density Mat{\'e}rn-like point processes are promising.  

The plot in \Cref{Materntest} shows the estimated pair correlation function $\hat g$ \citep[see][Section 4.3.3]{I} of the data and the pair correlation function $g$ both of a fitted Mat{\'e}rn I and a fitted Mat{\'e}rn II hard-core point process. 

Taking the minimum interpoint distance of $60.7\,\mu\mathrm{m}$ as an estimate for the hard-core distance $R$ (which is even a maximum likelihood estimate), fitting is here easily done by estimating the intensity $\lambda$ of the underlying Poisson process as the only remaining unknown parameter by the method of moments. That is, due to
\begin{equation*}
\lambda_I=\lambda\exp(-\lambda\,b_3\,R^3),\quad\lambda_{II}=\frac{1-\exp(-\lambda\,b_3\,R^3)}{b_3R^3},
\end{equation*}
for the intensities of the, respectively, Mat{\'e}rn I and Mat{\'e}rn II hard-core point process, an estimate of $\lambda$ can be obtained by solving for $\lambda$ in the equations $\hat{\lambda}=\lambda_I$ and $\hat{\lambda}=\lambda_{II}$, respectively, where $\hat{\lambda}=39.17\,\mathrm{mm}^{-3}$ is the empirical intensity of the data.
However, \Cref{Materntest} shows clear differences between the respective pair correlation functions indicating that none of the both Mat{\'e}rn hard-core point processes is a good model.

\begin{figure}
\begin{center}
\includegraphics[width=5.5cm]{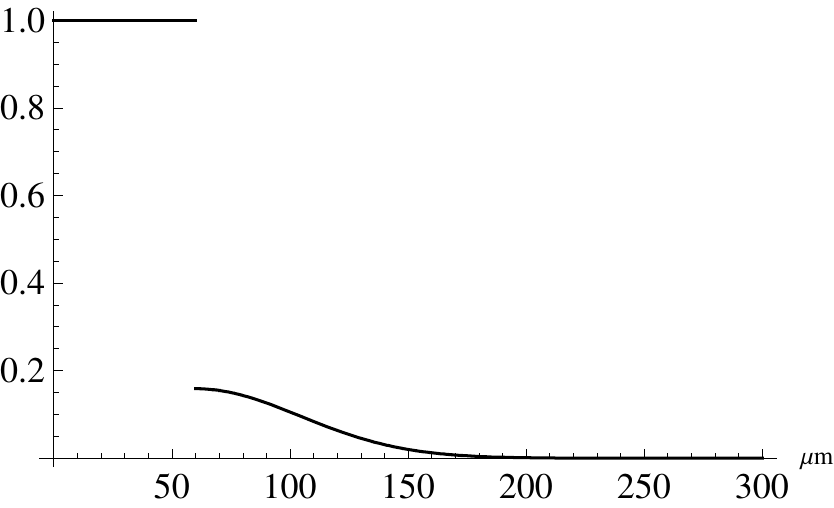}
\hspace{15mm}
\includegraphics[width=5.5cm]{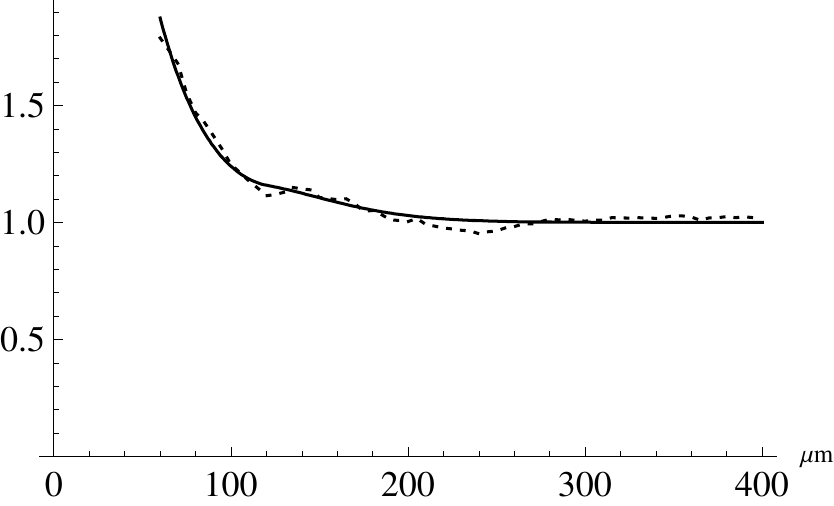}
\caption{Left, plot of $f_{R,a,b}$, $R=60.7\,\mu\mathrm{m}$, $a=6.3$ and $b=3917\,\mu\mathrm{m}^2$. Right, pair correlation functions $\hat g$ (dotted) of the network data and $g_{\thi}$ (solid line) of the fitted model.}
\label[figure]{fgstoyan}
\end{center}
\end{figure}

\begin{figure}
\begin{center}
\includegraphics[width=5.5cm]{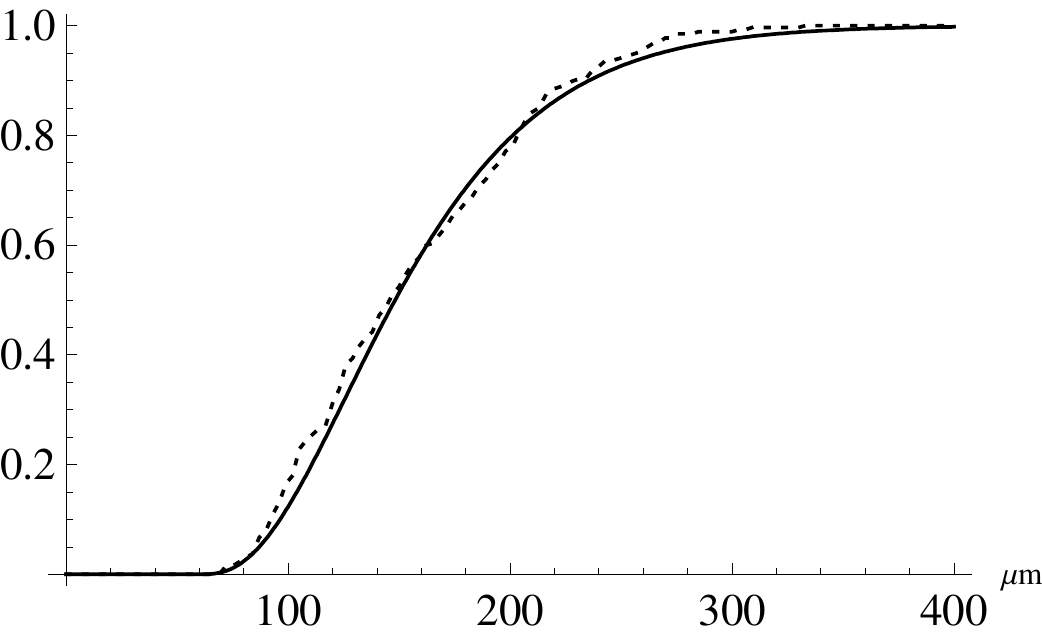}
\hspace{15mm}
\includegraphics[width=5.5cm]{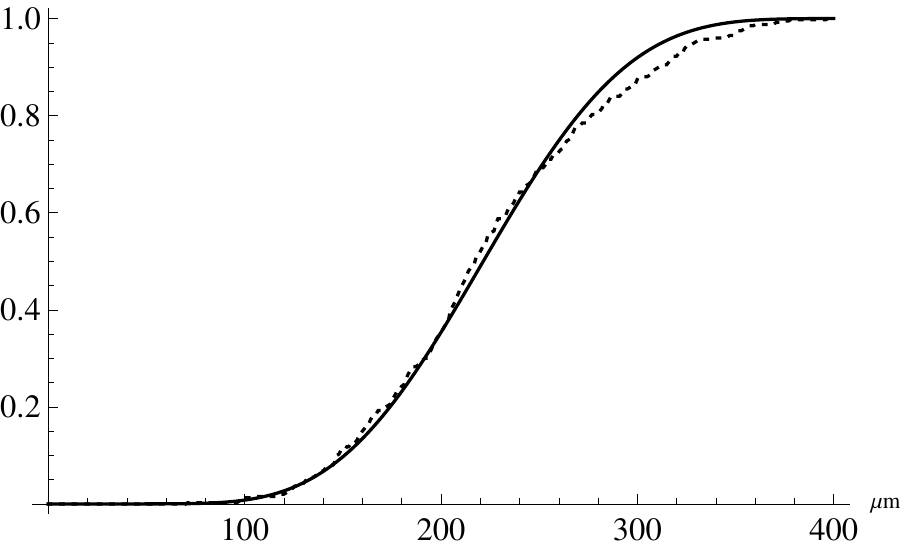}
\caption{Left, empirical nearest-neighbor distribution function of the network data (dotted) and nearest-neighbor distribution function of the fitted model (solid). Right, same for empirical empty space function.}
\label[figure]{neareststoyan}
\end{center}
\end{figure}

Since pure hard-core point process models of Mat{\'e}rn type seemed to be not appropriate we have then fitted a model $\mathrm{MatI}[\lambda,\po,f_{R,a,b}]$ from \Cref{model} with 
\begin{align*}
f_{R,a,b}(r)=\begin{cases}
    1 & 0 \leq r \leq R,\\
    \frac{1}{a}\,\exp\left(-\frac{(r-R)^2}{b}\right) & \text{otherwise,}
\end{cases}
\end{align*}  
with $R=60.7\,\mu\mathrm{m}$. Here, as parameter estimate the best possible choice of $(\lambda,\po,a,b)$ was taken, meant in the sense that under the condition $\lambda_{\thi}=\hat{\lambda}$ the contrast
\begin{equation}\label{eqn:mincontr}
\Delta(\lambda,\po,a,b)=\int_{r_{\min}}^{r_{\max}} [\hat g(r)-g_{\thi}(r)]^2 \dd r
\end{equation}
is minimized with respect to $(\lambda,\po,a,b)$, where $[r_{\min},r_{\max}]$ is a suitable domain. This is a variant of the well-known \emph{minimum contrast method} for parameter estimation \citep[][]{Di,He} where here the difference to be minimized depends on the pair correlation function as that summary statistics which is available at least via numerical integration for the models under consideration. The minimum contrast method using the pair correlation function has been also successfully applied by \citet{S1} and \citet[p. 183]{Mo}. This resulted in estimates
$\lambda=1919\,\mathrm{mm}^{-3}$, $\po=0.92$, $a=6.3$ and $b=3917\,\mu\mathrm{m}^2$. \Cref{fgstoyan} shows the estimated function $f_{R,a,b}$ as well as both the empirical pair correlation function $\hat{g}$ of the network data and the pair correlation function $g_{\thi}$ of the estimated model.
The visual finding from \Cref{fgstoyan} (right) that the fit is good can be suggested by formal tests. For instance, a \emph{deviation test} \citep[see][Section 7.4]{I} for the corresponding $L$-functions \citep[Section 4.3]{I} with global deviation measure
\begin{align*}
\Delta=\int_0^{\text{max}} |\hat L (r)-L_{\thi}(r)|^2 \dd r
\end{align*} ($p$-value $0.18$) with $k=99$ simulations indicates that now the model was chosen flexible enough to give a fit which mimics the second-order behavior of the data sufficiently well. Two other deviation tests with $k=99$ simulations using the nearest-neighbor distance distribution function as well as the empty space function (or 'spherical contact distribution function') \citep[][Section 4.2]{I} (see \Cref{neareststoyan} for plots) instead of the $L$-function then show that the fit is good also in other respects ($p$-values $0.14$ and $0.12$, respectively). 
%
\subsection{Patterns of deagglomerated alumina particles}
\begin{figure}
\begin{center}
\fbox{\includegraphics[width=6cm]{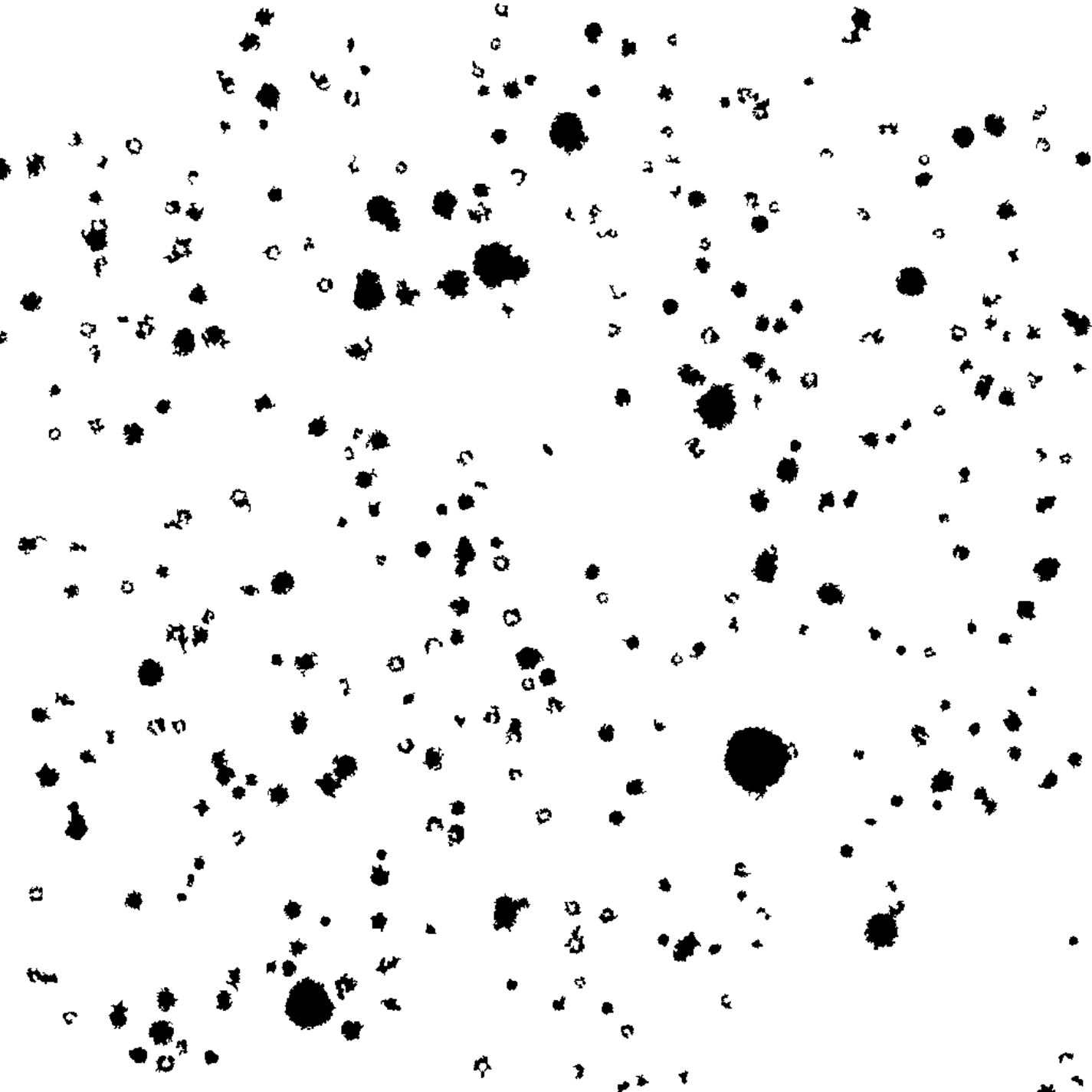}}
\caption{One of the used QICPIC images, size $1\,\mathrm{m}\times 1\,\mathrm{m}$.}
\label[figure]{Samples}
\end{center}
\end{figure}

The second data set are three samples of a mono-layer of deagglomerated alumina particles within water which serve as a starting point for the investigation of certain agglomeration processes not discussed here. The patterns, one shown in \Cref{Samples}, were obtained with a QICPIC sensor (Sympatec/Germany), which is a measurement device for dynamic picture analysis. For the test setup a liquid dispersing unit was used to get such a mono-layer flow of deagglomerated alumina particles through a flat cuvette where then the images were recorded. 
The median of the alumina particles is approximately $10\,\mu\mathrm{m}$. Due to the recording process, some of the particles look like open circles. Although they are all non-overlapping in space, some particles close together appear to be connected due to the projective nature of the recording.

The planar pattern of particles is quite sparse for which reason it might be modeled by a thinned marked Poisson process as in \Cref{modelmark,modelgeneral}. However, a Mat{\'e}rn-II-type model from \Cref{modelgeneral} might be comparatively more promising due to the higher attainable intensities, see \Cref{rmk:comparison}.

\begin{figure}
\begin{center}
\includegraphics[width=5.5cm]{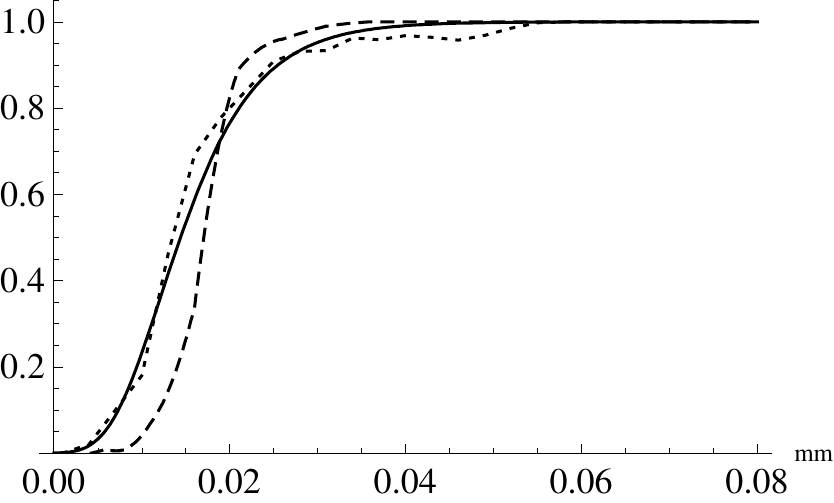}
\hspace{15mm}
\includegraphics[width=5.5cm]{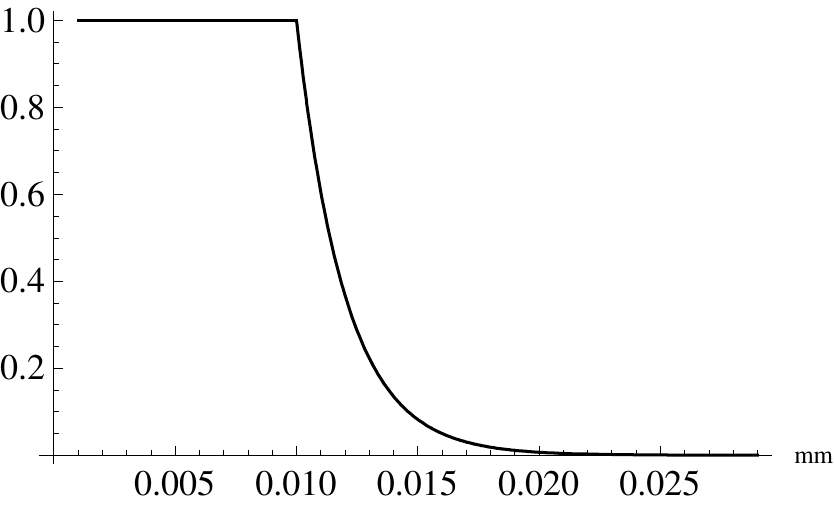}
\caption{Plot of $\hat g$ (dotted), $g_M$ (dashed) $g_{\thi}$ (solid), left. Function $f_c$ for fixed $c$ and $m,n$.}
\label[figure]{qicpcf}
\end{center}
\end{figure}

\begin{figure}
\begin{center}
\includegraphics[width=5.5cm]{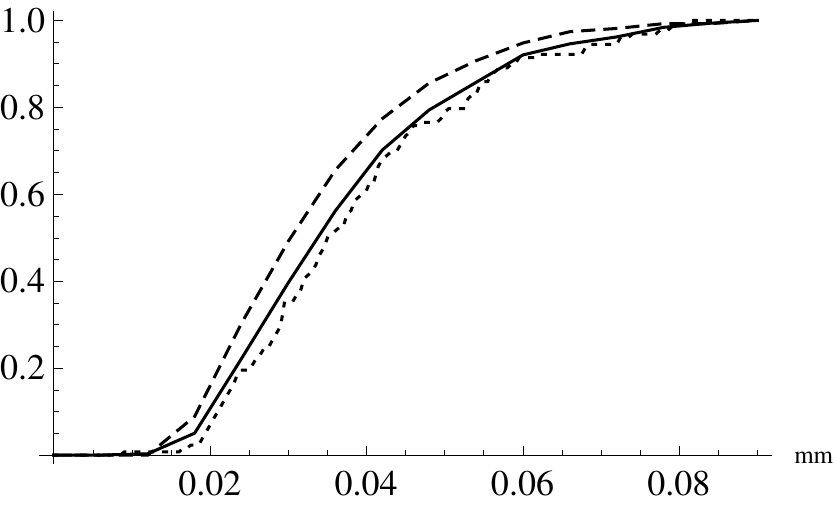}
\hspace{15mm}
\includegraphics[width=5.5cm]{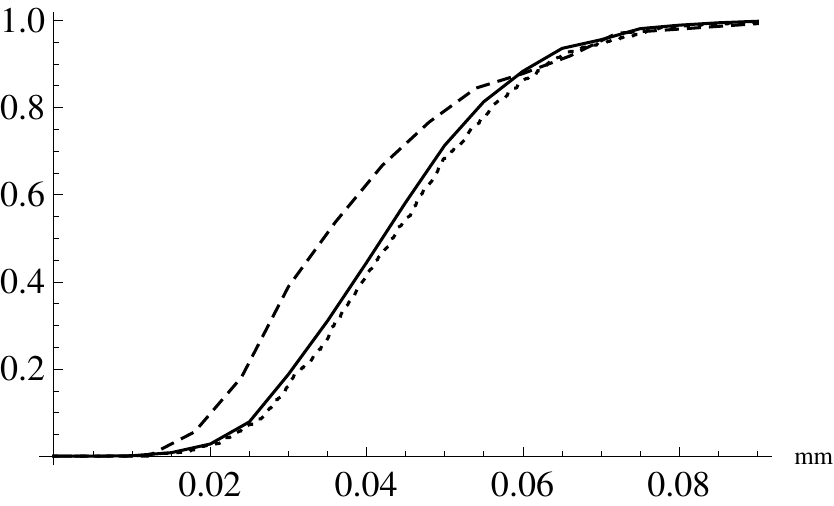}
\caption{Left, empirical c.d.f. of the nearest-neighbor distance of the data (dotted), Mat{\'e}rn (dashed) and $\Pi_{\thi}$ (solid). Right, same for empirical empty space function.}
\label[figure]{qiccheck}
\end{center}
\end{figure}

Our first attempt was to fit a Mat{\'e}rn II process for hard spheres as in \Cref{rmk:MIIspheres} with gamma-distributed radius marks, where for practical reasons the distribution was truncated at some high value. The distribution of the weight marks was chosen to be the uniform distribution on $[0,1]$. Here, three parameters, ($\lambda=315\,\mathrm{mm}^{-2}$, shape$=6.5$, rate$=0.00128\,\mathrm{mm}$), were estimated again by the minimum contrast method using the pair correlation function. A comparison of the resulting model pair correlation function $g_M$ and the empirical pair correlation function $\hat g$ of the data shown in \Cref{qicpcf} indicates that this kind of model is not flexible enough already for the second-order behavior of the data. 
This is supported also by the visual inspection of the corresponding ('Mat{\'e}rn') nearest-neighbor distance distribution function and empty space function, respectively, in \Cref{qiccheck}; see also \Cref{tabp} for several related deviation tests.

\begin{figure}
\begin{center}
\includegraphics[width=5.5cm]{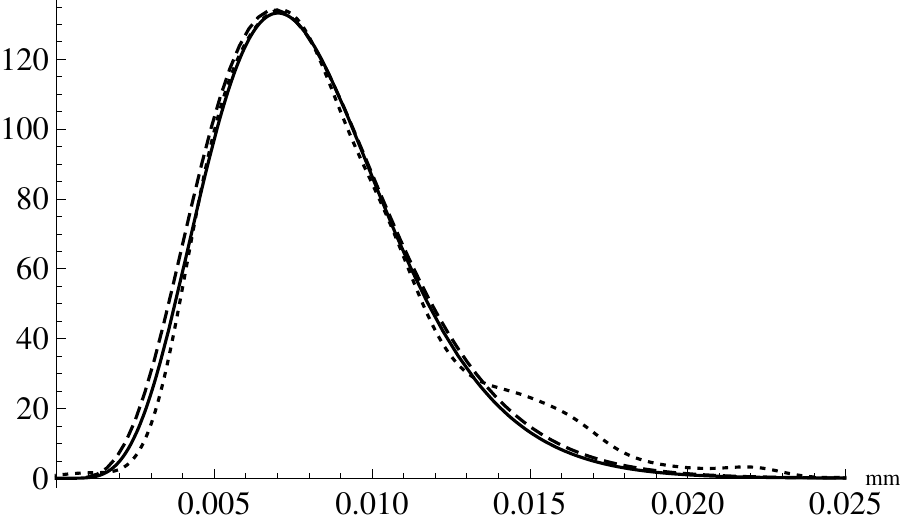}
\caption{Empirical p.d.f. of radii of the data (dotted), Mat{\'e}rn (dashed) and $\Pi_{\thi}$ (solid).}
\label[figure]{qicrad}
\end{center}
\end{figure} 
This motivated us to try modeling with the generalization introduced in \Cref{modelgeneral}. Here, the ansatz 
is $\Pi_{\thi} \sim \mathrm{MatII}[\lambda,\Gamma(\alpha,\beta),\text{Uniform}[0,1],\po,f_c]$ with
\begin{align*}
f_c(r,m,n)=
\begin{cases}
    1 & 0 \leq r \leq m+n\\
    \exp[-c(r-m-n)] & \text{otherwise}
\end{cases}
\end{align*}
(shown in \Cref{qicpcf}) and $\Gamma(\cdot,\cdot)$ as gamma distribution.

Parameter estimation by the minimum contrast method (\ref{eqn:mincontr}) with the pair correlation function yielded estimates $\lambda=335\,\mathrm{mm}^{-2}$, $\alpha=6.3$, $\beta=0.00127\,\mathrm{mm}$, $\po=0.96$ and $c=119\,\mathrm{mm}^{-1}$. 
The resulting characteristics like the pair correlation function $g_{\thi}$ shown in \Cref{qicpcf,qicrad,qiccheck} indicate a much better fit than the model of the first attempt. This is also supported by the corresponding deviation tests, see \Cref{tabp}, each based on $k=99$ simulations of the fitted model.
\begin{table}
\begin{center}
\begin{tabular}{ l | c | c | c | c}                     
 		       	& pcf  	& p.d.f radii & nearest-n. & empty sp.\\
\hline 
 Mat{\'e}rn  	& 0.00 	& 0.13        & 0.01 		& 0.00   \\
 $\Pi_{\thi}$   & 0.16 	& 0.17  	  & 0.07 		& 0.06    \\
\hline  
\end{tabular}
\caption{$p$-values for the deviation tests of the models 'Mat{\'e}rn' and $\Pi_{\thi}$ using the pair correlation function (pcf), the probability density function of the radius marks after thinning (p.d.f radii), the nearest neighbor distance distribution function (nearest-n.) and the empty space function (empty sp.), each based on $99$ simulations.}
\label[table]{tabp}
\end{center}
\end{table}
%
%
\section{Conclusions and outlook}
In this paper, we have examined a new class of point processes generalizing
the Mat{\'e}rn I and II hard-core point processes as well as the independent
thinning approach. Clearly, the proposed new model is not suited for very dense
and structured packings of particles, since, like for the Mat{\'e}rn processes, there is a relatively small upper bound
for the intensity for any given $f$ not vanishing almost everywhere. However, it provides a flexible and
simple to fit model in the class of dependent point process models.  Unlike for the also very flexible Gibbs point processes the simple mathematical structure allows for a simple and straight forward simulation and a direct computation of structure functions of the resulting point process, also simplifying the application of standard fitting procedures.  The approach allows for a interpretable descriptions of interactions, like interaction of shaped objects or non-deterministic death from competition. In our further research we also found that this class corresponds to distributions derived for snapshot point patterns in moving particle systems models we developed for the alumina particles. However, this relation has to be discussed in a separate article introducing these moving particle systems modeling.

Thus we think that this new class is worth considering for modeling various
real world point patterns and systems of particle centers.

%
\section{Proofs}
\subsection[Proof of Theorem 3]{Proof of \Cref{eq25}}\label{sec:proof3}
The intensity measure \citep{D,I,Mo,SK} $\alpha_{\thi}$ of $\Psi_{\thi}$ satisfies
\begin{align}\label{eqn:intRp}
\alpha_{\thi}(\mathrm{d}(x,m))=\po\,q_m\,\alpha_{\Psi}(\mathrm{d}(x,m))
\end{align}
where $\alpha_{\Psi}(\mathrm{d}(x,m))=\lambda\,\dd x\,\mu(\dd m)$ is the intensity measure of $\Psi$ and $q_m$ is the probability that a primary point with mark $m$ in $\Psi$ is retained as secondary event in $\Psi_{\thi}$ when thinning would be restricted to pairwise interaction, i.\,e., (\ref{eqn:rule1}) would be applied with $\po=1$.

Using Palm theory \citep{D,I,SK} and stationarity of $\Psi$, $q_m$ is the probability that under the reduced Palm distribution $P_{(o,m)}^!$ of $\Psi$, where $o$ denotes the origin, the point $(o,m)$ is not deleted by any other point. Since $\Psi$ is a marked Poisson point process this is, due to the Slivnyak-Mecke theorem \citep{I,Mo}, equivalent to the probability that the point $(o,m)$ is not deleted by any point from $\Psi$ when the same thinning rule is applied. 

Let $Q_m$ be the marked point process which consists of all points from $\Psi$ causing a deletion of $(o,m)$. Then $q_m$ is simply the probability that $Q_m$ has no points. Obviously, $Q_m$ is obtained by independent thinning of $\Psi$, i.\,e.
\begin{align*}
Q_m=\sum_{(x,l)\in\Psi} \beta(x,m,l)\,\delta_{(x,l)}\,,
\end{align*}
where $\beta(x,m,l)$ is Bernoulli-distributed with parameter $f(\|x\|,m,l)$ and $\delta_{(x,l)}$ denotes the Dirac measure centered on $(x,l)$. Hence, $Q_m$ is an inhomogeneous marked Poisson process \citep[Section 6.1]{I} with intensity measure
\begin{equation*}
\alpha_{Q_m}(\dd(x,l))=f(\|x\|,m,l)\,\alpha_{\Psi}(\dd(x,l)=f(\|x\|,m,l)\,\lambda\,\dd x\,\mu(\dd m).
\end{equation*}
Since $q_m$ is the void probability of the Poisson process $Q_m$ this implies
\begin{align}
q_m&=\PP(Q_m(\RR^d \times \RR)=0)=\exp\left(-\alpha_{Q_m}(\RR^d \times \RR)\right)\nonumber\\
&=\exp\left(-\lambda\int_{\RR}\int_{\RR^d}f(\|x\|,m,l)\,\dd x\,\mu(\dd l)\right)\nonumber\\
&=\exp\left(-\lambda d\,b_d\int_{\RR}\int_0^{\infty}f(r,m,l)r^{d-1}\,\dd r\,\mu(\dd l)\right)\label{eq22a}
\end{align}
using polar coordinates in the last step. Hence, due to \Cref{eqn:intRp}, the point process of unmarked points of $\Psi_{\thi}$ has intensity
\begin{align*}
\lambda_{\thi}
&=\lambda\,\po\int_{\RR}q_m\,\mu(\dd m)\\
&=\lambda\,\po\int_{\RR}\exp\left(-\lambda d\,b_d\int_{\RR}\int_0^{\infty}f(r,m,l)r^{d-1}\,\dd r\,\mu(\dd l)\right)\,\mu(\dd m).
\end{align*}
%
%
\subsection[Proof of Theorem 4]{Proof of \Cref{eq31}}\label{sec:proof4}
Let $\kappa_{m,n}(r)$ be the probability that two points in $\Psi$ with mark $m$ and $n$ a distance $r$ apart are both retained in $\Psi_{\thi}$ when (\ref{eqn:rule2}) is applied with $\po=1$. Then the second-order factorial moment measure \citep{D,I,Mo,SK} $\alpha_{\thi}^{(2)}$ of $\Psi_{\thi}$ satisfies
\begin{equation}\label{eqn:secRp}
\alpha_{\thi}^{(2)}(\dd(x,m,y,n))=\po^2\kappa_{m,n}(\|x-y\|)\,\alpha^{(2)}_{\Psi}(\dd(x,m,y,n)),
\end{equation}
where the second-order factorial moment measure $\alpha^{(2)}_{\Psi}$ of $\Psi$ factorizes to
\begin{equation}\label{eqn:secPsi}
\alpha^{(2)}_{\Psi}(\dd(x,m,y,n))=\alpha_{\Psi}(\dd(x,m))\,\alpha_{\Psi}(\dd(y,n))=\lambda^2\,\dd x\,\mu(\dd m)\,\dd y\,\mu(\dd n)
\end{equation}
since $\Psi$ is a Poisson process \citep[see][]{D}. Using again Palm theory and the Slivnyak-Mecke theorem, $\kappa_{m,n}(r)$ equals the probability that the two points $(o,m)$ and $(z,n)$, $\|z\|=r$, do not delete each other and are non of them is deleted by any point from $\Psi$ according to the thinning rule (\ref{eqn:rule2}) with $\po=1$. Since both events are independent and the probability of the first event is $(1-f(r,m,n))^2$, it follows
\begin{align}
\kappa_{m,n}(r)=(1-f(r,m,n))^2\,\PP(W_{z,m,n}(\RR^d \times \RR)=0),\label{eq27}
\end{align}
where $W_{z,m,n}$ is the marked point process which consists of all points from $\Psi$ causing a deletion of $(o,m)$ \emph{or} $(z,n)$. Due to independent thinning, i.\,e.,
\begin{align*}
W_{z,m,n}=\sum_{(x,l)\in\Psi} \max\{\gamma_0(x,m,l),\gamma_r(x,n,l)\} \delta_{(x,l)}\,,
\end{align*}
where $\gamma_r(x,a,b)$ is Bernoulli-distributed with parameter $f(\|x-r\cdot v\|,a,b)$, $\|v\|=1$, $W_{z,m,n}$ is an inhomogeneous marked Poisson process with intensity measure
\begin{align*}
\alpha_{W_{z,m,n}}(\dd(x,l))&=[f(\|x\|,m,l)+f(\|x-r\cdot v\|,n,l)-f(\|x\|,m,l)f(\|x-r\cdot v\|,n,l)]\\
&\quad\times\lambda\,\dd x\,\mu(\dd l).
\end{align*}
According to \Cref{eq27}, this yields
\begin{align}
\kappa_{m,n}(r)&= (1-f(r,m,n))^2\exp\left(-\alpha_{W_{z,m,n}}(\RR^d\times\RR)\right)\notag \\
&= (1-f(r,m,n))^2\exp\left(-\lambda\,\int_{\RR}\int_{\RR^d} f(\|x\|,m,l)\,\dd x\,\mu(\dd l)\right) \notag\\
&\qquad\times\exp\left(-\lambda\,\int_{\RR}\int_{\RR^d} f(\|x\|,n,l)\,\dd x\,\mu(\dd l)\right) \notag \\
&\qquad\times\exp\left(\lambda\,\int_{\RR}\int_{\RR^d} f(\|x\|,m,l)f(\|x-r\cdot v\|,n,l)\,\dd x\,\mu(\dd l)\right) \notag\\
&=(1-f(r,m,n))^2q_m\,q_n \exp\left(\lambda\,\int_{\RR} f(\cdot,m,l) \astrosun f(\cdot ,n,l)(r)\,\mu(\dd l) \right)\,,\label{eq29}
\end{align}
using \Cref{eq22a} and the radial convolution \Cref{eqn:selfconv} in the last step. Abbreviating the last factor of the product in \Cref{eq29} by $q_{m,n}(r)$ and combining equations \Cref{eqn:secRp} and \Cref{eqn:secPsi}, the second-order product density of the point process $\tilde{\Psi}_{\thi}$ of unmarked points of $\Psi_{\thi}$ is
\begin{align*}
\rho_{\thi}(r)
&=\lambda^2\,\po^2\!\int_{\RR}\int_{\RR}\!(1\!-\!f(r,m,n))^2\,q_m\,q_n\,q_{m,n}(r)\mu(\dd m)\mu(\dd n).
\end{align*}
Due to $g_{\thi}(r)=\rho_{\thi}(r)/\lambda_{\thi}^2$ for $r\geq 0$ \citep{I} and \Cref{eq25} this yields the asserted form of the pair correlation function $g_{\thi}$, i.\,e., in particular, $\po^2$ cancels out.
%
%
\subsection[Proof of Theorem 5]{Proof of \Cref{eq43}}
Basically, the idea of the proof is the same as in Section \ref{sec:proof3}. Here, let $q_m(w)$ be the probability that the point $(o,m,w)$ is not deleted by any point from $\Pi$ when the thinning rule (\ref{eqn:rule3}) with $\po=1$ is applied. Then $\po\,q_m(w)$ is the density of the intensity measure of $\Pi_{\thi}$ with respect to the intensity measure $\alpha_{\Pi}$ of $\Pi$, $\alpha_{\Pi}(\dd(x,m,w))=\lambda\,\dd x\,\nu_m(\dd w)\,\mu(\dd m)$, and $q_m(w)$ equals the probability that the marked point process $Q_{m,w}$ consisting of all points from $\Pi$ causing a deletion of $(o,m,w)$ is empty. Since $Q_{m,w}$ is a Poisson process with intensity measure $\mathbbm{1}\{v\leq w\}f(\|x\|,m,l)\,\alpha_{\Pi}(\dd(x,l,v))$ this yields
\begin{align}
q_m(w)&=\PP(Q_{m,w}(\RR^d \times\RR\times\RR)=0)\nonumber\\
&=\exp\left(-\lambda\int_{\RR}\int_{\RR}\int_{\RR^d}\mathbbm{1}\{v\leq w\}f(\|x\|,m,l)\,\dd x\,\nu_l(\dd v)\,\mu(\dd l)\right)\nonumber\\
&=\exp\left(-\lambda d\,b_d\int_{\RR}\int_0^{\infty}F_{\nu_l}(w)\,f(r,m,l)r^{d-1}\,\dd r\,\mu(\dd l)\right)\label{eq41}
\end{align}
and, finally,
\begin{equation*}
\lambda_{\thi}=\lambda\,\po\int_{\RR}\int_{\RR}q_m(w)\,\nu_m(\dd w)\,\mu(\dd m).
\end{equation*} 
%
%
\subsection[Proof of Theorem 6]{Proof of \Cref{eq49}}
The main arguments of the proof of \Cref{eq31} in \Cref{sec:proof4} can be carried over. Let $\po^2\kappa_{m,n}(w,t,r)$ be the probability that two points in $\Pi$ with marks $m$ and $n$ and weights $w$ and $t$ a distance $r$ apart are both retained in $\Pi_{\thi}$. Then $\kappa_{m,n}(w,t,r)$ equals the probability that (a) the two points $(o,m,w)$ and $(z,n,t)$, $\|z\|=r$, do not delete each other and (b) non of them is deleted by any point from $\Pi$ according to the thinning rule (\ref{eqn:rule3}) with $\po=1$. Again, both events are independent, and the probability of event (a) is
\begin{equation*}
1-[\PP(A)+\PP(B)-\PP(A\cap B)]=1-f(r,m,n)+\mathbbm{1}\{t=w\}[f(r,m,n)^2-f(r,m,n)]
\end{equation*}
since the probabilities that (A) $(z,n,t)$ deletes $(o,m,w)$, that (B) $(o,m,w)$ deletes $(z,n,t)$, and that both delete each other are $\mathbbm{1}\{t\geq w\}f(r,m,n)$, $\mathbbm{1}\{t\leq w\}f(r,m,n)$, and $\mathbbm{1}\{t=w\}f(r,m,n)^2$, respectively. The probability of event (b) is the probability that the Poisson process $W_{z,m,n,w,t}$ consisting of all points of $\Pi$ causing a deletion of $(o,m,w)$ or $(z,n,t)$ is empty. Hence, using $q_m(w)$ from (\ref{eq41}) as shorthand, it equals
\begin{align*}
\PP(W_{z,m,n,w,t}(\RR^d\times\RR\times\RR)=0)=q_m(w)\,q_n(t)\,q_{m,n}(w,t,r),
\end{align*}
where
\begin{align*}
q_{m,n}(w,t,r)=\exp\left(\lambda\int_{\RR}\int_{\RR^d} F_{\nu_l}(\min\{w,t\})\,f(\|x\|,m,l)f(\|x-r\cdot v\|,n,l)\,\dd x\,\mu(\dd l)\right),
\end{align*}
since $W_{z,m,n,w,t}$ has intensity measure
\begin{align*}
&\big[\mathbbm{1}\{u\leq w\}f(\|x\|,m,l)+\mathbbm{1}\{u\leq t\}f(\|x\|,n,l)\\
&\qquad-\mathbbm{1}\{u\leq w\}f(\|x\|,m,l)\mathbbm{1}\{u\leq t\}f(\|x\|,n,l)\big]\,\alpha_{\Pi}(\dd(x,l,u))
\end{align*}
and
\begin{equation*}
\int_{\RR}\mathbbm{1}\{u\leq w\}\mathbbm{1}\{u\leq t\}\nu_l(\dd u)=F_{\nu_l}(\min\{w,t\}).
\end{equation*}
Therefore, the second-order product density of $\tilde{\Pi}_{\thi}$ is
\begin{align*}
\rho_{\thi}(r)
&=\lambda^2\,\po^2\int_{\RR}\int_{\RR}\int_{\RR}\int_{\RR}\kappa_{m,n}(w,t,r)\,\nu_m(\dd w)\,\nu_n(\dd t)\,\mu(\dd m)\,\mu(\dd n)\\
&=\lambda^2\po^2\!\int_\RR \int_\RR (1-f(r,m,n))\,I_r(m,n)\,\mu(\dd m)\, \mu(\dd n)
\end{align*}
where
\begin{align*}
I_r(m,n)= \int_\RR \int_\RR q_m(w)\,q_n(t)\,q_{m,n}(w,t,r)\,\nu_m(\dd w)\,\nu_n(\dd t)\,.
\end{align*}
Note that the summand $\mathbbm{1}\{t=w\}(f(r,m,n)^2-f(r,m,n))$ in the first factor of the integrand has been left out since its integral vanishes due to the assumed continuity of the distributions $\nu_m$ and $\nu_n$.
\section*{Acknowledgments}
The authors would like to thank the German Science Foundation (DFG) for supporting the scientific work within the framework of the Collaborative Research Centre ``Multi-Functional Filters for Metal Melt Filtration - A Contribute towards Zero Defect Materials'' (SFB 920). They are very grateful to J.~Fritzsche and F.~Heuzeroth for providing their particle data sets and to D.~Stoyan for inspiring discussions on the topic.
%


\bibliographystyle{unsrtnat}     
\bibliography{materngen} 

\begin{thebibliography}{25}
\providecommand{\natexlab}[1]{#1}
\providecommand{\url}[1]{\texttt{#1}}
\expandafter\ifx\csname urlstyle\endcsname\relax
  \providecommand{\doi}[1]{doi: #1}\else
  \providecommand{\doi}{doi: \begingroup \urlstyle{rm}\Url}\fi

\bibitem[Stoyan et~al.(1995)Stoyan, Kendall, and Mecke]{SK}
D.~Stoyan, W.~S. Kendall, and J.~Mecke.
\newblock \emph{Stochastic Geometry and its Applications}.
\newblock Wiley, Chichester, 2nd edition, 1995.

\bibitem[Mat{\'e}rn(1960)]{M60}
B.~Mat{\'e}rn.
\newblock Spatial variation. stochastic models and their application to some
  problems in forest surveys and other sampling investigations.
\newblock \emph{Medd. Statens Skogsforskningsinst.}, 49\penalty0 (5):\penalty0
  1--144, 1960.

\bibitem[Mat{\'e}rn(1986)]{M}
B.~Mat{\'e}rn.
\newblock \emph{Spatial Variation}.
\newblock Lecture Notes in Statistics 36, Springer, New York, 1986.

\bibitem[Picard et~al.(2005)Picard, Kouyate, and Dessard]{PKD}
N.~Picard, M.~Kouyate, and H.~Dessard.
\newblock Tree density estimations using a distance method in {M}ali savanna.
\newblock \emph{Forest Science}, 51\penalty0 (1):\penalty0 7--18, 2005.

\bibitem[Stoyan(1987)]{S4}
D.~Stoyan.
\newblock Statistical analysis of spatial point processes: a soft-core model
  and cross correlations of marks.
\newblock \emph{Biometrical J.}, 29:\penalty0 971--980, 1987.

\bibitem[Warren(1972)]{W}
W.~G. Warren.
\newblock Point processes in forestry.
\newblock In P.~S.~W. Lewis, editor, \emph{Stochastic Point Processes}, pages
  801--816. Wiley, New York, 1972.

\bibitem[Baccelli and B{\l}aszczyszyn(2009)]{BB}
F.~Baccelli and B.~B{\l}aszczyszyn.
\newblock \emph{Stochastic Geometry and Wireless Networks, Volume II ---
  Applications}, volume 4, No 1--2 of \emph{Foundations and Trends in
  Networking}.
\newblock NoW Publishers, 2009.

\bibitem[Busson and Chelius(2009)]{BC}
A.~Busson and G.~Chelius.
\newblock Point processes for interference modeling in {CSMA/CA} ad hoc
  networks.
\newblock In \emph{Sixth ACM International Symposium on Performance Evaluation
  of Wireless Ad Hoc, Sensor, and Ubiquitous Networks (PE-WASUN 09)}, 2009.

\bibitem[Haenggi(2011)]{H1}
M.~Haenggi.
\newblock Mean interference in hard-core wireless networks.
\newblock \emph{IEEE Communications Letters}, 15\penalty0 (8):\penalty0
  792--794, 2011.

\bibitem[Stoyan(1988)]{S3}
D.~Stoyan.
\newblock Thinnings of point processes and their use in the statistical
  analysis of a settlement pattern with deserted tillages.
\newblock \emph{Statistics}, 19:\penalty0 45--56, 1988.

\bibitem[Illian et~al.(2008)Illian, Penttinen, Stoyan, and Stoyan]{I}
J.~Illian, A.~Penttinen, H.~Stoyan, and D.~Stoyan.
\newblock \emph{Statistical Analysis and Modelling of Spatial Point Patterns}.
\newblock Wiley, Chichester, 2008.

\bibitem[Batista and Maguire(1998)]{BM}
J.~L. Batista and A.~D. Maguire.
\newblock Modeling the spatial structure of tropical forests.
\newblock \emph{Forest Ecology and Mangement}, 110:\penalty0 293--314, 1998.

\bibitem[Stoyan and Stoyan(1985)]{S2}
D.~Stoyan and H.~Stoyan.
\newblock On one of {M}at{\'e}rns hard-core point process models.
\newblock \emph{Math. Nachr.}, 122:\penalty0 205--214, 1985.

\bibitem[M{\aa}nsson and Rudemo(2002)]{Ma}
M.~M{\aa}nsson and M.~Rudemo.
\newblock Random patterns of nonoverlapping convex grains.
\newblock \emph{Adv. appl. Prob.}, 34:\penalty0 718--738, 2002.

\bibitem[Daley and Vere-Jones(2008)]{D1}
D.~J. Daley and D.~Vere-Jones.
\newblock \emph{An Introduction to the Theory of Point Processes, Vol. II}.
\newblock Springer, New York, 2008.

\bibitem[Daley and Vere-Jones(1988)]{D}
D.~J. Daley and D.~Vere-Jones.
\newblock \emph{An Introduction to the Theory of Point Processes}.
\newblock Springer, New York, 1988.

\bibitem[Strauss(1975)]{Str}
D.~J. Strauss.
\newblock A model for clustering.
\newblock \emph{Biometrika}, 62:\penalty0 467--475, 1975.

\bibitem[M{\o}ller and Waagepetersen(2003)]{Mo}
J.~M{\o}ller and R.~P. Waagepetersen.
\newblock \emph{Statistical Inference and Simulation for Spatial Point
  Processes}.
\newblock Chapman \& Hall/CRC, Boca Raton, 2003.

\bibitem[Gavrikov and Stoyan(1995)]{Ga}
V.~L. Gavrikov and D.~Stoyan.
\newblock The use of marked point processes in ecological and environmental
  forest studies.
\newblock \emph{Environ. Ecolog. Statist.}, 2:\penalty0 331--344, 1995.

\bibitem[Andersson et~al.(2006)Andersson, H{\"a}ggstr{\"o}m, and
  M{\aa}nsson]{A}
J.~Andersson, O.~H{\"a}ggstr{\"o}m, and M.~M{\aa}nsson.
\newblock The volume fraction of a non-overlapping germ-grain model.
\newblock \emph{Electronic Communications in Probability}, 11:\penalty0 78--88,
  2006.

\bibitem[Sok et~al.(2002)Sok, Knackstedt, Sheppard, Pinczewski, Lindquist,
  Venkatarangan, and Paterson]{So}
R.~M. Sok, M.~A. Knackstedt, A.~P. Sheppard, W.~V. Pinczewski, W.~B. Lindquist,
  A.~Venkatarangan, and L.~Paterson.
\newblock Direct and stochastic generation of network models from tomographic
  images: effect of topology on two phase flow properties.
\newblock \emph{Trans. Porous Media}, 46:\penalty0 345--371, 2002.

\bibitem[Tscheschel and Stoyan(2003)]{Ts}
A.~Tscheschel and D.~Stoyan.
\newblock On the estimation variance for the specific {E}uler-{P}oincar{\'e}
  characteristic of random networks.
\newblock \emph{J. Microsc.}, 211:\penalty0 80--88, 2003.

\bibitem[Diggle(2003)]{Di}
P.~J. Diggle.
\newblock \emph{Statistical Analysis of Spatial Point Patterns}.
\newblock Arnold, London, 2003.

\bibitem[Heinrich(1992)]{He}
L.~Heinrich.
\newblock Minimum contrast estimates for parameters of spatial ergodic point
  processes.
\newblock In \emph{Transactions of the 11th Prague Conference on Random
  Processes, Information Theory and Statistical Decision Functions}, pages
  479--492. Academic Publishing House, 1992.

\bibitem[Stoyan and Stoyan(1996)]{S1}
D.~Stoyan and H.~Stoyan.
\newblock Estimating pair correlation functions of planar cluster processes.
\newblock \emph{Biometrical J.}, 38:\penalty0 259--271, 1996.

\end{thebibliography}
\end{document}